\documentclass{PoS}
\usepackage[authoryear]{natbib}
\bibpunct{(}{)}{;}{a}{}{,}

\newcommand{\eq}[1]{\begin{equation}  #1 \end{equation}}
\newcommand{\ba}[1]{\left\langle #1 \right\rangle}
\newcommand{\br}[1]{\left( #1 \right)}

\title{Weak gravitational lensing with the Square Kilometre Array}

\ShortTitle{Weak lensing with the Square Kilometre Array}

\author{\speaker{M.~L.~Brown},$^1$ {D.~J.~Bacon},$^{2}$
  {S.~Camera},$^1$ {I.~Harrison},$^1$ {B.~Joachimi},$^3$
  {R.~B.~Metcalf},$^4$ {A.~Pourtsidou},$^2$ {K.~Takahashi},$^5$
  {J.~A.~Zuntz},$^{1}$ {F.~B.~Abdalla},$^{3,6}$ {S.~Bridle},$^1$
  {M.~Jarvis},$^7$ {T.~D.~Kitching},$^3$ {L.~Miller},$^7$ {P.~Patel}$^8$ \\
$^1$Jodrell Bank Centre for Astrophysics, School of Physics and
  Astronomy, The University of Manchester, Oxford Road, Manchester M13 9PL, UK\\
$^2$Institute of Cosmology and Gravitation, University of Portsmouth,
  Burnaby Road, Portsmouth PO1 3FX, UK\\
$^3$Department of Physics and Astronomy, University College London, Gower Street, London, WC1E 6BT, UK\\
$^4$Dipartimento di Fisica e Astronomia, Universit\'{a} di Bologna, viale B. Pichat 6/2 , 40127, Bologna, Italy\\
$^5$Faculty of Science, Kumamoto University, 2-39-1 Kurokami, Kumamoto 860-8555, Japan\\
$^6$Department of Physics and Electronics, Rhodes University, PO Box 94, Grahamstown, 6140, South Africa \\
$^7$Astrophysics, Department of Physics, University of Oxford, Oxford OX1 3RH, UK\\
$^8$Department of Physics, University of Western Cape, Cape Town 7535, South Africa\\
\\
Email: \email{m.l.brown@manchester.ac.uk}}

\abstract{We investigate the capabilities of various stages of the SKA
to perform world-leading weak gravitational lensing surveys. We
outline a way forward to develop the tools needed for pursuing weak
lensing in the radio band. We identify the key analysis challenges
and the key pathfinder experiments that will allow us to address
them in the run up to the SKA. We identify and summarize the unique
and potentially very powerful aspects of radio weak lensing surveys,
facilitated by the SKA, that can solve major challenges in the field
of weak lensing. These include the use of polarization and rotational
velocity information to control intrinsic alignments, and the new
area of weak lensing using intensity mapping experiments. We show how
the SKA lensing surveys will both complement and enhance corresponding
efforts in the optical wavebands through cross-correlation techniques
and by way of extending the reach of weak lensing to high redshift.}

\FullConference{
Advancing Astrophysics with the Square Kilometre Array\\
June 8-13, 2014\\
Giardini Naxos, Italy}

\begin{document}

\section{Background}

\subsection{Cosmology with weak lensing surveys}
Weak gravitational lensing is the coherent distortion in the shapes of
distant galaxies due to the deflection of light rays by intervening
mass distributions. Measurements of the effect on large scales is
termed ``cosmic shear'' and has emerged as a powerful probe of
late-time cosmology over the last 15 years
(see e.g.~\citealt{heymans13} for recent results from the CFHTLenS
survey). Since gravitational lensing is sensitive to
the \emph{total} (i.e. dark plus baryonic) matter content of the
Universe, it has great potential as a very robust cosmological probe,
to a large degree insensitive to the complications of galaxy formation
and galaxy bias. One of the most promising aspects of weak lensing
measurements is their combination with redshift information: such
measurements are then a sensitive probe of both the geometry of the
Universe and of the evolution of structure over the course of cosmic
time. In turn, these latter effects are dependent on the nature of the
dominant dark energy component in the Universe and/or on modifications
to the theory of General Relativity on large scales.

The observed distortions in the shapes of distant galaxies yields an
estimate of the lensing shear field, $\gamma$. Since gravity is a
potential theory, the shear at angular position $\theta$ can be
related to a lensing potential ($\psi$) as
\begin{equation}
\gamma_{ij}(\theta) = \left( \delta_i \delta_j -\frac{1}{2} \delta^K_{ij}
\delta^2 \right) \psi(\theta),
\label{eq:shear_def}
\end{equation}
where $\delta_i \equiv r (\delta_{ij} - \hat{r}_i \hat{r}_j \nabla_i)$
is a dimensionless, transverse differential operator, and $\delta^2 =
\delta_i \delta^j$ is the transverse Laplacian. The indices $(i, j)$
each take the values $(1, 2)$. In equation~(\ref{eq:shear_def}) we have
assumed a flat sky which is an excellent approximation for the scales
of interest (i.e. from $\sim100$ $h^{-1}$kpc to $\sim100$
$h^{-1}$Mpc). The lensing potential can in turn be related to the 3-d
gravitational potential, $\Phi(\mathbf{r})$ by
(e.g.~\citealt{kaiser98, hu00}) 
\begin{equation}
\psi(\theta) = \frac{2}{c^2} \int^r_0 dr' \left(\frac{r-r'}{rr'}\right)\Phi(\mathbf{r'}),
\end{equation}
where $r$ is the comoving distance to the sources. 

In the limit of weak lensing ($\gamma << 1$), and in the absence of
intrinsic correlations in galaxy ellipticities, one can form an
unbiased estimator for the shear field at a given sky position from
the average of the observed galaxy ellipticities at that
position. Since the shear field induced by large scale strucutre is
small (typically $\sim$ a few $\%$) compared to the intrinsic
dispersion in galaxy ellipticities ($\sigma_\epsilon \sim 0.3$) one
needs to average over many background galaxies in order to obtain a
precise measurement. Measurements of cosmic shear can be affected by a
number of instrumental and astrophysical systematic effects. The
primary insturmental effect of concern is anisotropies in the point
spread function (or beam) of the telescope which can mimic a cosmic
shear signal. On the astrophysical side, alignments in the
\emph{intrinsic} shapes of galaxies can also mimic a cosmic shear
signal. Consequently a great deal of effort is currently focused on
careful instrument design for the next generation of weak lensing
surveys as well as theoretical modelling and numerical simulations of
intrinsic alignment effects.

Observationally, to date the field of weak lensing has largely been the preserve of
optical surveys due to the much larger number densities of background
galaxies achieved in such surveys. However, this will change with the
advent of the Square Kilometre Array which will reach number densities
of well-detected and well-resolved galaxies of up to $\sim5$ galaxies
arcmin$^{-2}$ over several thousand deg$^2$ in Phase 1, and $\sim10$
galaxies arcmin$^{-2}$ over $3\pi$ steradians in Phase 2. In addition,
as described in this chapter, the radio offers truly unique approaches
to measuring weak lensing that are (i) not available to optical
surveys and (ii) potentially extremely powerful in minimizing the most
worrying systematic effects in weak lensing cosmology.

SKA surveys will also extend the reach of weak lensing beyond that of
the optical mega-surveys of LSST and Euclid. First the typical
redshifts probed by SKA surveys will go beyond that of optical
surveys. Secondly the SKA offers the prospects of measuring the weak
lensing distortion in 21cm HI intensity maps. Such an intensity
mapping lensing survey will not only extend to the high-redshifts
inaccessible to optical surveys but will also include precise redshift
information to accompany the distortion signal. This signal will fill
the gap between traditional galaxy lensing signal (where the sources
are typically located at $z\sim1$) and the CMB lensing signal
originating at $z=3000$. It thus holds the promise of yielding very
precise cosmological constraints on the evolution of structure during
early times where structures are better described with linear physics
and can provide unique insight into so-called ``early dark energy''
models which exhibit observable signatures at early times.

\subsection{Radio weak lensing studies to date}
To date, the application of weak lensing analyses to radio data
has been rather limited due to the relatively low number
density of background galaxies achieved in large scale radio
surveys. The only major weak lensing analysis of a radio survey was
performed in \cite{chang04} who detected a cosmic shear signal in
the Faint Images of the Radio Sky at Twenty cms (FIRST) survey 
\citep{becker95}, conducted with the Very Large Array (VLA). Although
the number density of sources in FIRST was low by optical
standards (FIRST contains $\sim 90$ sources deg$^{-2}$ compared with
typically $\sim10$ sources arcmin$^{-2}$ in deep optical lensing
surveys), a detection of cosmic shear on large scales was achieved by
virtue of the large survey area covered ($\sim$ 10,000 deg$^2$). 

The analysis of \cite{chang04} made use of the shapelets method
\citep{chang02, refregier03} to measure the shapes of the FIRST sources
directly from the $uv$ interferometric data and included a thorough
treatment of the possible systematics that can affect radio-based
lensing shear estimates in particular.  

More recently, \cite{patel10} revisited the idea of radio-based
lensing shear estimation through a re-analysis of the combined VLA +
MERLIN observations of the Hubble Deep Field North
\citep{muxlow05}. This work presented an implementation of
image-based shapelets for shear estimation and also highlighted the
advantages of cross-correlating radio-based and optical-based shear
estimates obtained over the same area of sky. 

\subsection{The path to SKA weak lensing}
A great deal of algorithm development and new analysis techniques are
required in order to develop the field of radio weak lensing beyond
the initial studies mentioned above. Key areas where work is required
are the development of shape and/or shear estimation techniques that
are well suited to radio interferometer datasets, the development of
cross-correlation techniques for combining optical and radio-based
shear estimates and the development and demonstration of novel
radio-specific lensing approaches making use of polarization and
rotational velocity information. A number of SKA pathfinder and
precursor telescopes lend themselves naturally to developing several
of these areas.

A key pathfinder telescope for demonstrating the long-baseline high
resolution observations suited for weak lensing is the e-MERLIN
interferometer based in the UK. A number of e-MERLIN legacy programs
are well-suited to pathfinding the techniques for SKA weak lensing. Of
particular note are the
e-MERGE\footnote{http://www.e-merlin.ac.uk/legacy/projects/emerge.html}
and
SuperCLASS\footnote{http://www.e-merlin.ac.uk/legacy/projects/superclass.html}
projects. e-MERGE is a multi-tiered project that (amongst other
things) will image the GOODS-N field to 0.5 $\mu$Jy rms in the central
100 arcmin$^2$ and to 1 $\mu$Jy rms in the surrounding 800
arcmin$^2$. The SuperCLASS project complements e-MERGE. Its primary
science driver is to detect the weak lensing effects of a supercluster
of galaxies located at $z=0.2$. SuperCLASS will image a 1.75 deg$^2$
field to 4 $\mu$Jy rms. Both e-MERGE and SuperCLASS will perform their
primary observations at 1.4~GHz with a resolution of 200 mas. Both
will therefore act as training experiments for demonstrating shape
measurement and shear extraction algorithms on high-resolution radio
data, on the path to the SKA.

A further opportunity exists to refine and test radio lensing analyses
with the upgraded JVLA. A series of sky surveys
(VLASS\footnote{https://science.nrao.edu/science/surveys/vlass}) 
are currently being discussed for the JVLA and these will potentially
include a deep fields component. If conducted with an array
configuration that includes long baselines, these could be extremely
powerful for pathfinding weak lensing techniques in the radio over
larger survey areas, approaching 10 deg$^2$ \citep{brown13}.

Other opportunities for demonstrating weak lensing techniques in the
run up to the SKA include LOFAR
surveys\footnote{http://lofar.strw.leidenuniv.nl} 
(when the international baselines are included) and the 
CHILES\footnote{http://www.mpia-hd.mpg.de/homes/kreckel/CHILES/index.html} 
and CHILES con
Pol\footnote{http://www.aoc.nrao.edu/$\sim$chales/chilesconpol/} 
surveys on the JVLA. The LOFAR surveys will be critical
for testing weak lensing techniques at low frequencies while the
CHILES and CHILES con Pol surveys will prove useful for testing novel
ideas for radio lensing including the use of polarization information
and HI rotational velocity measurements (see
Section~\ref{sec:pol_vrot_techniques}).

\section{Cosmic shear with radio continuum surveys}

\subsection{Accessing the largest scales and the highest redshifts}
\begin{figure}[t!]
\centering
\includegraphics[width=15cm]{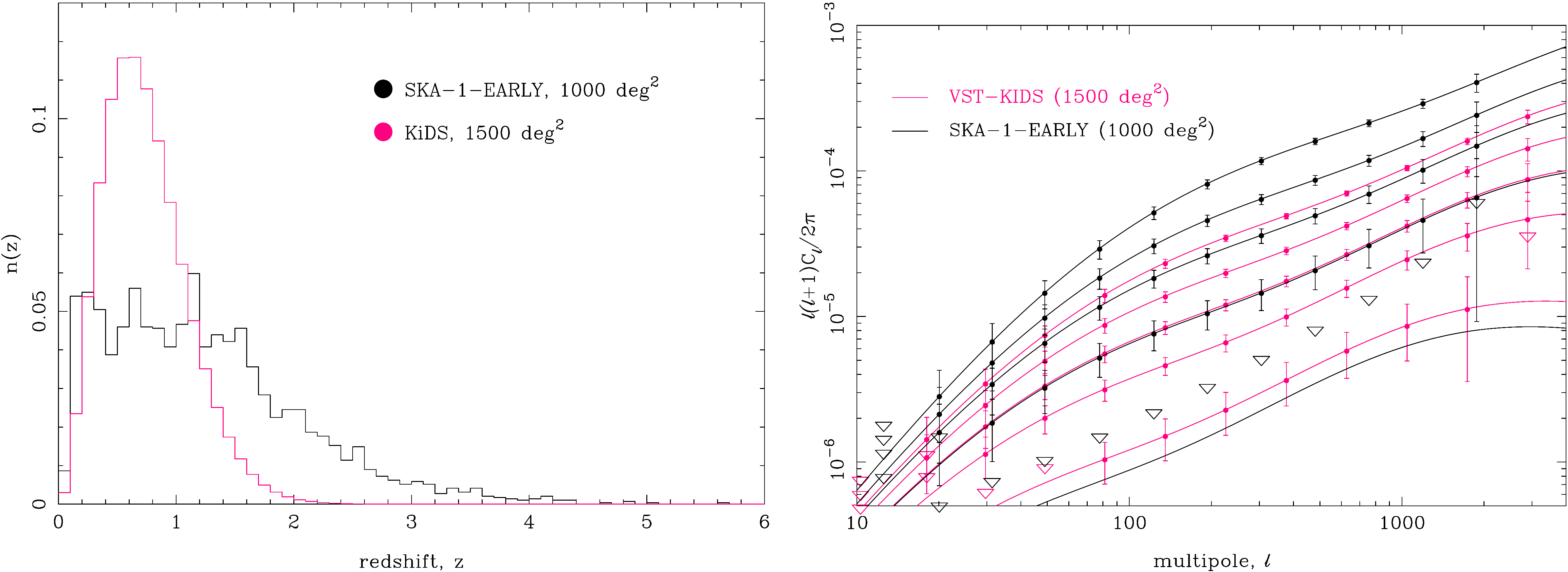}
\vspace{0mm}
\caption{\emph{Left panel:} The redshift distribution of source galaxies for
  a 1000 deg$^2$ weak lensing survey requiring 2 years observing
  time on the SKA1-early facility. Also shown is the redshift
  distribution for the 1500 deg$^2$ VST-KiDS optical lensing
  survey. The $n(z)$ extends to higher redshifts in the radio survey
  and probes a greater range of cosmic history. \emph{Right panel:}
  The corresponding constraints on a 5-bin tomographic power spectrum
  analysis. For both experiments, we assumed an RMS dispersion in
  ellipticity measurements of $\gamma_{\rm rms} = 0.3$ and the
  tomographic bins have been chosen such that the bins are populated
  with equal numbers of galaxies. Note how the radio survey extends to
  higher redshifts where the lensing signal is stronger and therefore
  easier to measure. Open triangles denote $1\sigma$
  upper limits on a bandpower. Note that only the auto power
  spectra in each bin are displayed though much cosmological
  information will also be encoded in the cross-correlation spectra
  between the different $z$-bins.}
\label{fig:ska1early_cls}
\end{figure}

\begin{figure}[t!]
\centering
\includegraphics[width=15cm]{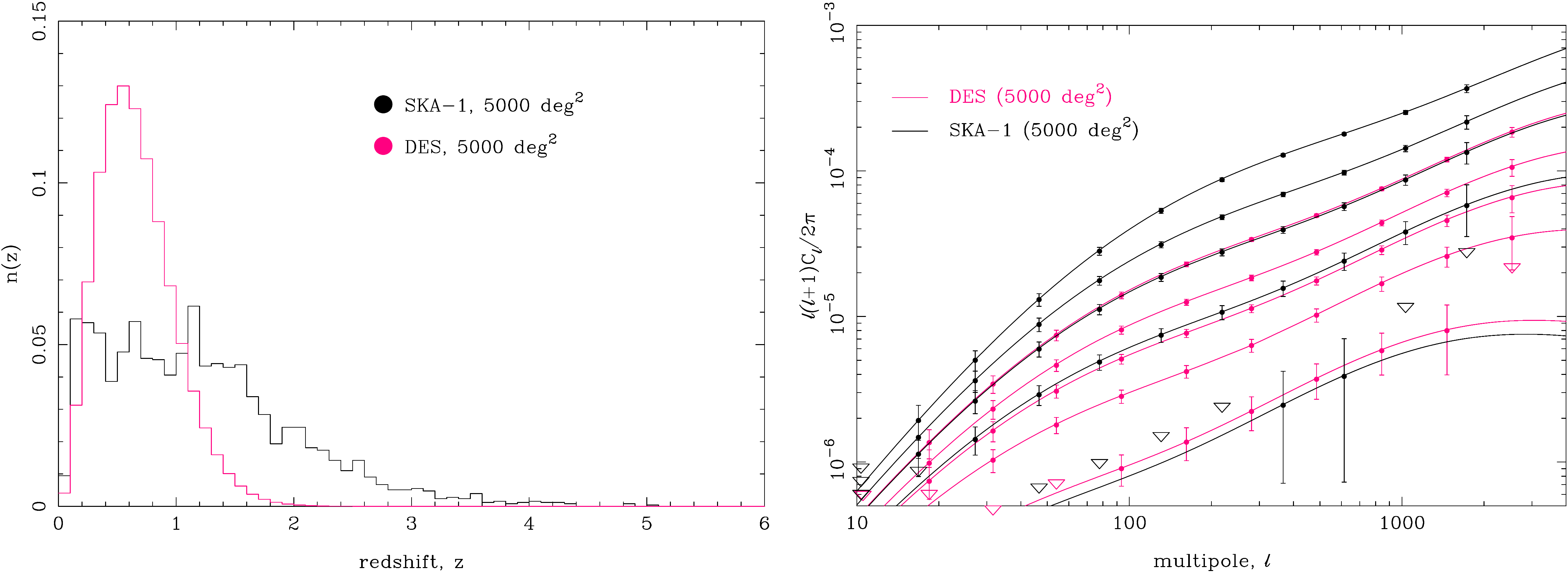}
\vspace{0mm}
\caption{As Fig.~\protect\ref{fig:ska1early_cls} but for a 5000 deg$^2$ weak
  lensing survey requiring 2 years observing time on the full SKA1
  facility. Also shown for comparison are the $n(z)$ distribution and
  forecasted power spectrum constraints for the 5000 deg$^2$ Dark
  Energy Survey.}
\label{fig:ska1_cls}
\end{figure}

\begin{figure}[t!]
\centering
\includegraphics[width=15cm]{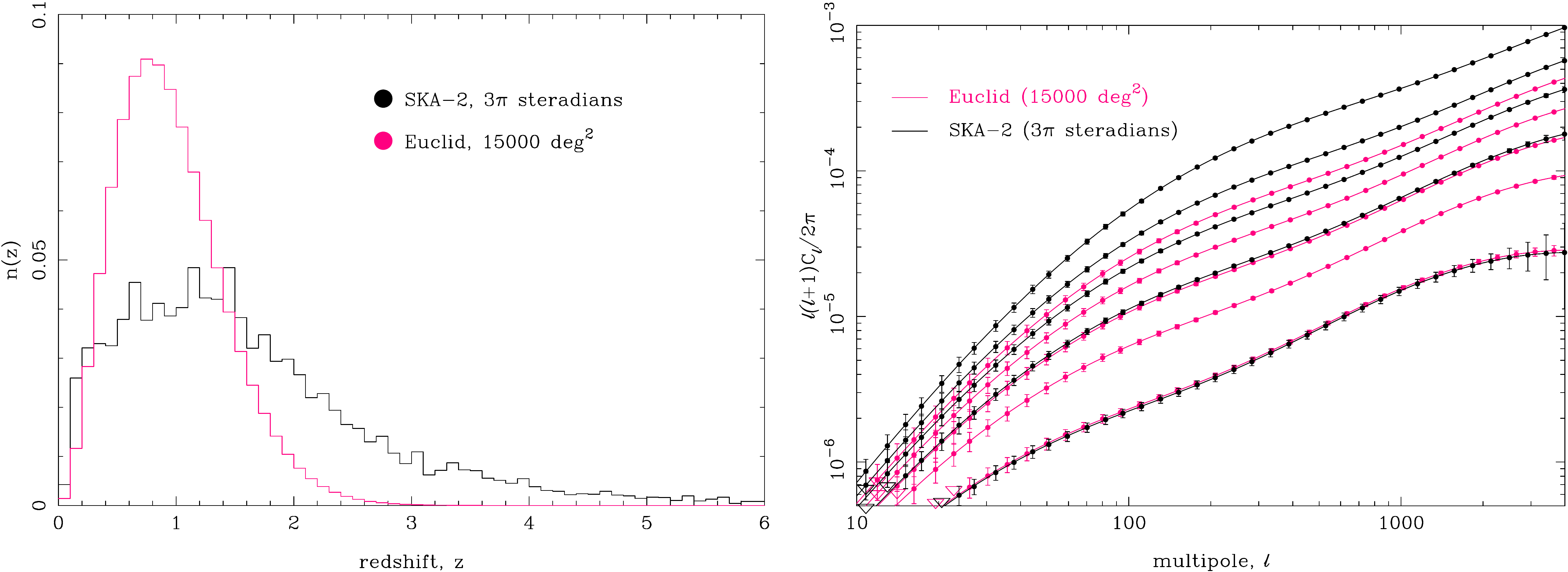}
\vspace{0mm}
\caption{As Fig.~\protect\ref{fig:ska1early_cls} but for a $3\pi$
  steradian weak lensing survey requiring 2 years observing time on
  SKA2. Also shown for comparison are the $n(z)$ distribution
  and forecasted power spectrum constraints for the 15000 deg$^2$
  Euclid satellite mission.}
\label{fig:ska2_cls}
\end{figure}

One of the potentially most exciting aspects of weak lensing surveys
with the SKA will be the extra reach in terms of accessing the largest
scales in the Universe and going beyond the reach of other weak
lensing surveys in terms of the redshifts that are probed. We
illustrate this in Figs.~\ref{fig:ska1early_cls}--\ref{fig:ska2_cls}
where the redshift distributions of weak lensing source galaxies and
the corresponding forecasted errors on a set of tomographic shear
power spectra are presented. We present these forecasts for a 2-year
continuum survey with the SKA at three different stages of development
-- an early phase of SKA1 comprising 50\% of the envisaged sensitivity
targeting a survey area of 1000 deg$^2$; the full SKA1 surveying 5000
deg$^2$, and a SKA2 survey covering $3\pi$ steradians. To generate
these forecasts, we have adopted the performance specifications for
SKA1 outlined in \cite{braun13}. In particular, we have modeled the
performance of Band 2 of the SKA-Mid facility as under the current
baseline design, this telescope and frequency band combination
provides the most powerful survey speed for the high angular
resolution observations required for weak lensing. We have further
assumed an object detection threshold of $S/N > 10$ and an angular
resolution requirement of $\theta_{\rm res} = 0.5$ arcsec. To model
the redshift and flux dependence of the source population, we have
made use of the SKA Design Studies (SKADS) simulations of
\cite{wilman08}, updated to match the galaxy number counts observed in
the deepest radio surveys performed to date \citep{muxlow05,
  morrison10, schinnerer10}. We have also assumed an RMS dispersion in
intrinsic galaxy ellipticities of $\gamma_{\rm rms} = 0.3$.

For comparison on these plots, we also show the corresponding
forecasts for optical weak lensing surveys that will be conducted over
similar survey areas and on comparable timescales. Specifically, we
consider the VST-KiDS \citep{deJong13}, the Dark Energy
Survey\footnote{http://www.darkenergysurvey.org} and the Euclid
satellite mission \citep{laureijs11}. The observational parameters
adopted to produce these forecasts are summarized in
Table~\ref{tab:tab1}. 
\begin{table}
\centering
\caption{Observational parameters used to produce the power spectrum
 and cosmological parameter forecasts in
 Figs.~\protect\ref{fig:ska1early_cls}--\protect\ref{fig:params}.}
\label{tab:tab1}
\vspace{5mm}
\begin{tabular}{lccc}
\hline
Survey & $A_{sky}$ (deg$^2$) & $n_{gal}$ (arcmin$^{-2}$) & $z_{m}$ \\
\hline
\hline
SKA1-early & 1000  & 3.0 & 1.0  \\
VST-KiDS   & 1500  & 7.5 & 0.6  \\
\hline
SKA1       & 5000  & 2.7   & 1.0 \\
DES        & 5000  & 6.0   & 0.6 \\
\hline
SKA2       & 30940 & 10    & 1.6 \\
Euclid     & 15000 & 30    & 0.9 \\
\hline
\end{tabular}
\end{table}
In all cases, the radio surveys extend to higher redshift than the
corresponding optical probes. They thus hold the potential to probe
the power spectrum at higher redshift providing a more sensitive lever
arm with which to constrain the growth of structure over cosmic time.

\begin{figure*}
\begin{center}
\includegraphics[width=0.4\linewidth, trim=0.2cm 0.0cm 1cm 1cm, clip=true]{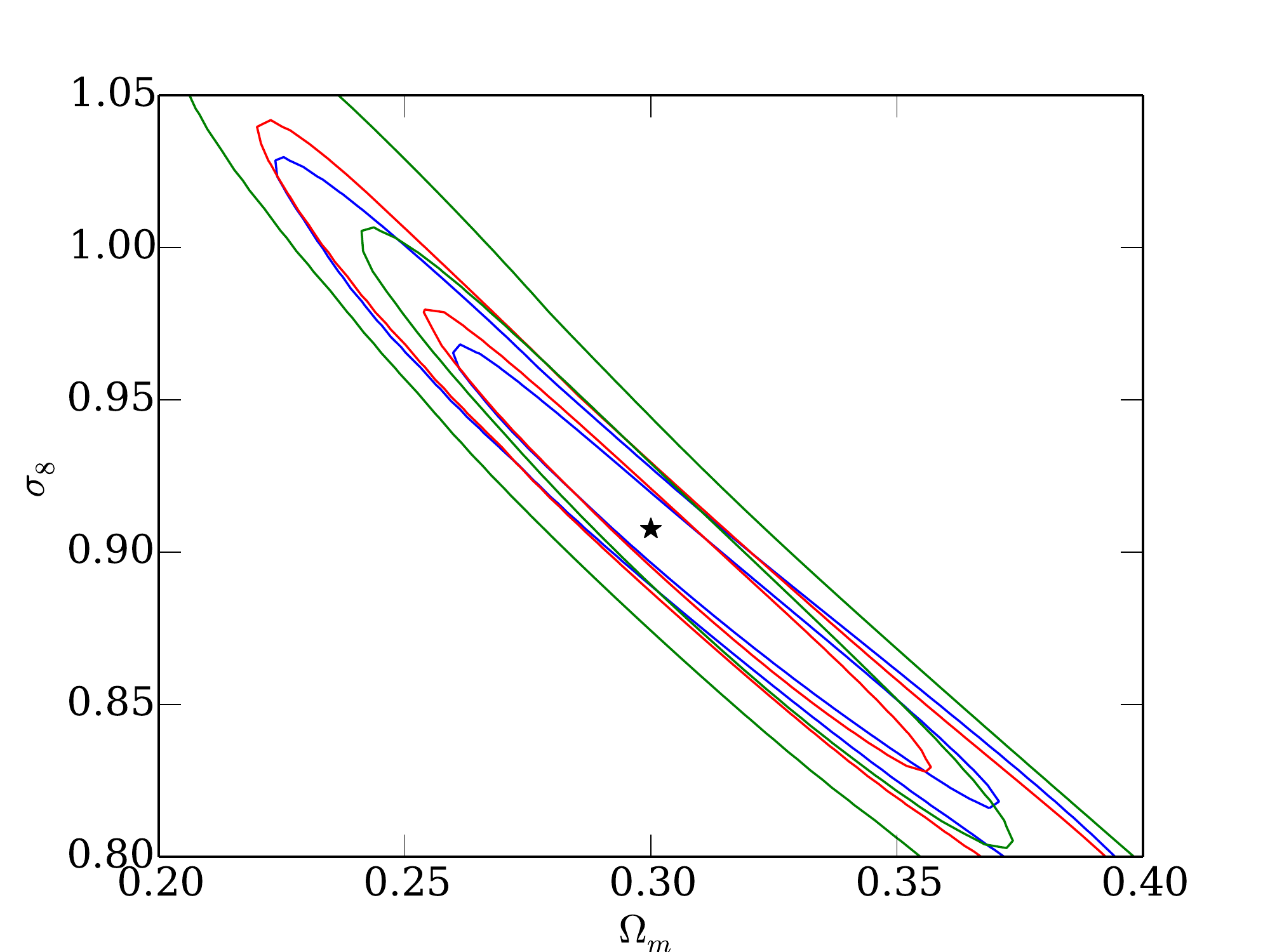}  
\includegraphics[width=0.4\linewidth, trim=0.2cm 0.0cm 1cm 1cm, clip=true]{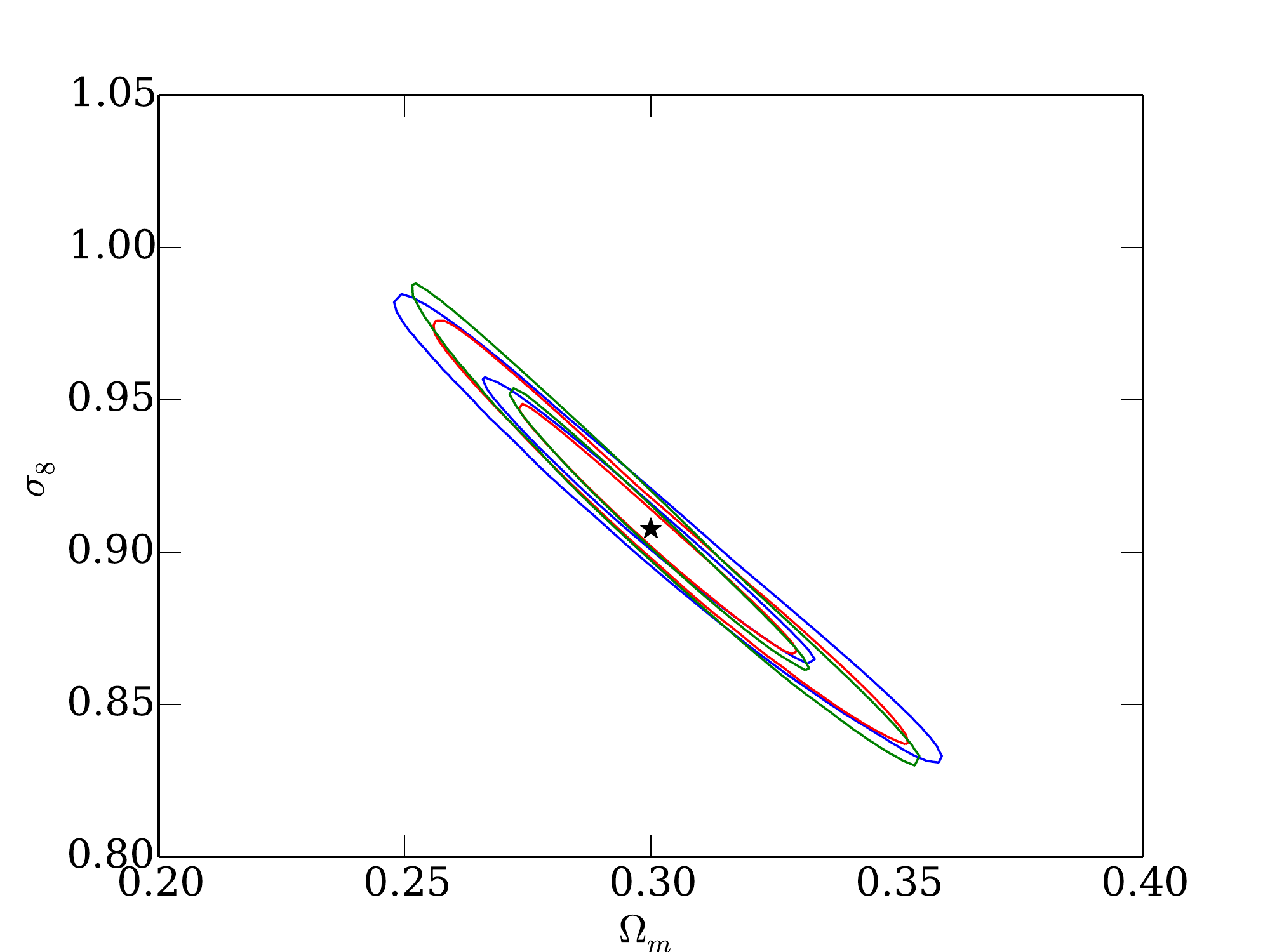}\\
\vspace{5mm}
\includegraphics[width=0.4\linewidth, trim=0.2cm 0.0cm 1cm 1cm, clip=true]{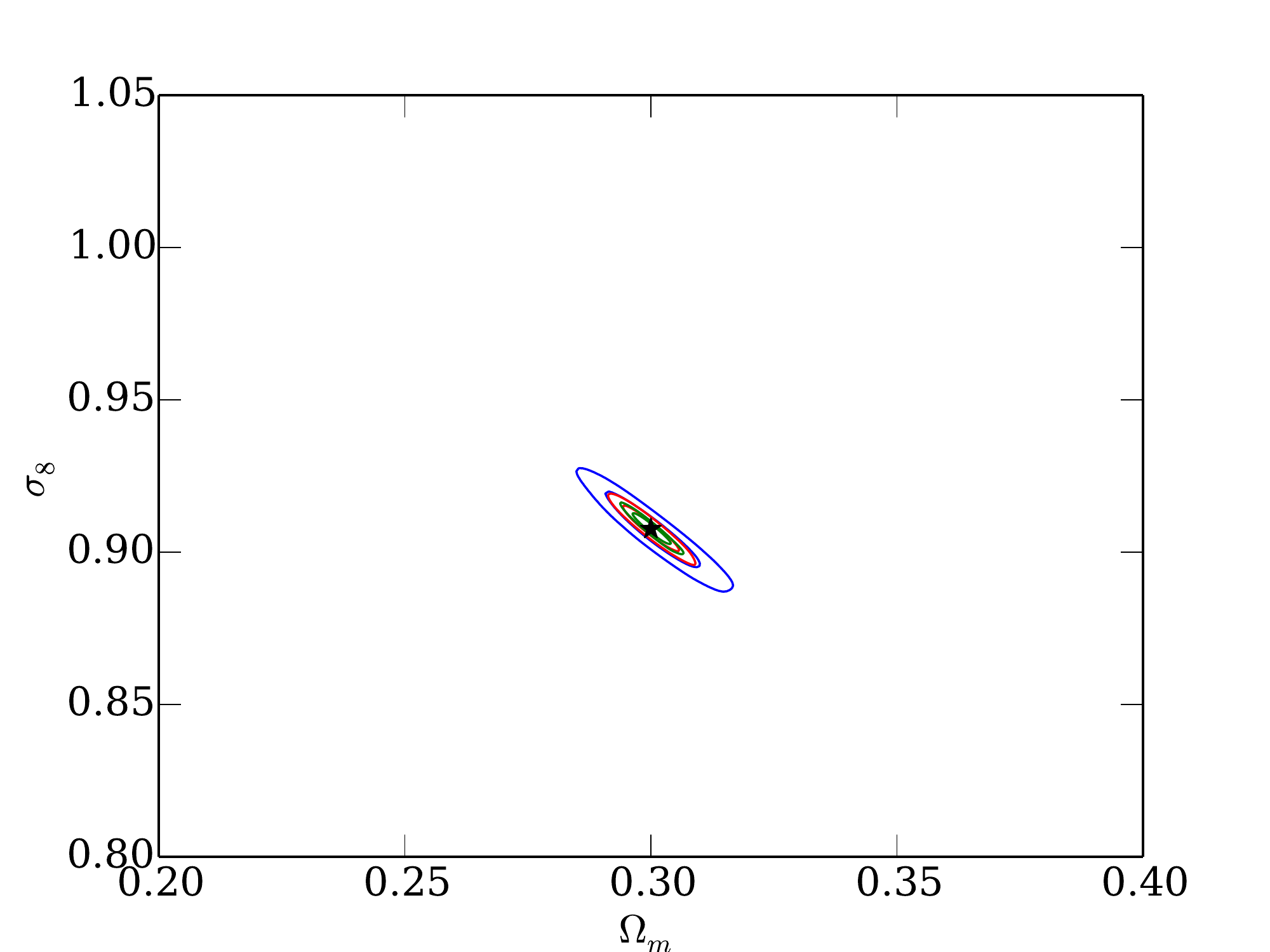}\\
\caption{Forecasted constraints in the $\sigma_8 - \Omega_m$ parameter
  space for 2-year continuum surveys using SKA-Mid/Band 2 with
  SKA1-early (\emph{upper left panel}), SKA1 (\emph{upper right panel})
  and SKA2 (\emph{lower panel}) performance parameters. For each of
  these cases,  we present the constraints for survey areas of $1000$
  deg$^2$ (blue) $5000$ deg$^2$(red) and $3\pi$ steradians
  (green). \label{fig:params}}
\end{center}
\end{figure*}

Fig.~\ref{fig:params} presents forecasted constraints on the matter
density ($\Omega_m$) and matter power spectrum normalization
($\sigma_8$) cosmological parameters for the same envisaged SKA
surveys as were adopted to generate
Figs.~\ref{fig:ska1early_cls}--\ref{fig:ska2_cls}. Note that these
forecasts are presented for the case of the standard 6-parameter
$\Lambda$CDM model and no prior information is assumed -- that is the
projected constraints are coming solely from the envisaged SKA weak
lensing survey.

To generate the constraints, we have computed a simple shear power
spectrum covariance matrix from \cite{TakadaJain}, and we use the {\sc
  CosmoSIS} cosmological parameter estimation code \citep{zuntz14} to
compute power spectra, parameter constraints, and marginalised
contours. Note that no systematic errors are included in the analysis;
errors are purely statistical. However, we have attempted to take into
account anticipated knowledge and uncertainties regarding photometric
and spectroscopic redshift estimates for the background galaxy
population. For SKA1-early, we have assumed that we have no
spectroscopic redshift information and that we have photo-$z$
estimates from overlapping optical surveys with errors $\sigma_z =
0.05 (1+z)$ up to a limiting redshift of 1.5. To model the much larger
uncertainties expected for the high-$z$ radio galaxies, we adopt
$\sigma_z = 0.3 (1+z)$ so that a $z = 2$ galaxy has a redshift
uncertainty of $\pm \sim1$. For SKA1, we additionally assume that we
will have spectroscopic redshifts from overlapping HI observations for
15\% of the $z < 0.6$ population. Finally for SKA2, we assume we have
spectroscopic redshifts for 50\% of the $z < 2.0$ population.  The
forecasts presented in Fig.~\ref{fig:params} account for these
redshift uncertainties.

We see from Fig.~\ref{fig:params} that even the SKA1-early survey
targeting the smallest sky area can provide competitive constraints
on cosmological parameters --- the forecasted constraints for the 1000
deg$^2$ SKA1-early survey are a factor of $\sim$5 better than the
tomographic weak lensing analysis of the current state-of-the-art
CFHTLenS data \citep{heymans13}. We also see large improvements in the
constraints obtainable with each subsequent stage of the SKA --- the
constraints obtainable with SKA1 are broadly comparable with the KiDS
and DES optical surveys while SKA2 is competitive with
Euclid. 

Fig.~\ref{fig:params} also demonstrates that our nominal choice of target
survey areas for the three stages of the SKA are, broadly speaking,
optimal choices from the point of view of constraining these cosmological
parameters --- for SKA1-early the 1000 deg$^2$ survey provides
the strongest constraints, for SKA1 the 5000 deg$^2$ survey performs
the best while for SKA2, the $3\pi$ steradian survey provides
the best constraints. 

\subsection{The promise of radio observations to suppress weak lensing systematics} 
Optical and radio surveys, such as Euclid and/or LSST and the SKA,
have a particularly useful synergy in reducing and quantifying the
impact of systematic effects which may dominate each survey alone on
some scales. By cross-correlating the shear estimators from one of
these surveys with those of the other, several systematic errors are
mitigated.

We can see this by writing the contributions to an optical ($o$) or
radio ($r$) shear estimator:
\begin{eqnarray}
\gamma^{(o)}&=&\gamma_{\rm grav} + \gamma^{(o)}_{\rm int} + \gamma^{(o)}_{\rm sys}\\
\gamma^{(r)}&=&\gamma_{\rm grav} + \gamma^{(r)}_{\rm int} + \gamma^{(r)}_{\rm sys},
\end{eqnarray}
where $\gamma_{\rm grav}$ is the gravitational shear we are seeking,
$\gamma_{\rm int}$ is the intrinsic ellipticity of the object, and
$\gamma_{\rm sys}$ are systematic errors induced by the telescope. If
we correlate optical shears with optical shears, or radio shears with
radio shears, we obtain terms like
\begin{equation}
\langle \gamma \gamma \rangle = \langle \gamma_{\rm grav} \gamma_{\rm
  grav} \rangle + \langle \gamma_{\rm grav} \gamma_{\rm int} \rangle +
\langle \gamma_{\rm int} \gamma_{\rm int} \rangle + \langle
\gamma_{\rm sys} \gamma_{\rm sys} \rangle,
\end{equation}
where the first term is the gravitational signal we seek, the second
term is the GI intrinsic alignment \citep{hirata04}, the third term is
the II intrinsic alignment (e.g.~\citealt{heavens00}), and the final term
is the contribution from systematics. All of these terms could be
similar size on certain scales, which is damaging to cosmological
constraints. On the other hand, if we cross-correlate the optical
shears with radio shears, we obtain
\begin{equation}
\langle \gamma^{(o)} \gamma^{(r)} \rangle = \langle \gamma_{\rm grav}
\gamma_{\rm grav} \rangle + \langle \gamma_{\rm grav}
\gamma^{(o)}_{\rm int} \rangle + \langle \gamma_{\rm grav}
\gamma^{(r)}_{\rm int} \rangle + \langle \gamma^{(o)}_{\rm int}
\gamma^{(r)}_{\rm int} \rangle + \langle \gamma^{(o)}_{\rm sys}
\gamma^{(r)}_{\rm sys} \rangle.
\end{equation}
The second and third terms are the GI alignment \citep{hirata04}, which
still survives. However, the fourth term involves the correlation
between optical and radio shapes, which will be less than that between
one frequency alone as the emission mechanisms are different
(c.f.~\citealt{patel10} where no correlation at zero lag was found). This
term is therefore reduced. Most importantly, the fifth term involving
systematics is expected to be zero, as the systematics in these two
telescopes, which are of completely different design and function, are
not expected to be correlated at all. We are therefore able to remove
the dangerous systematics correlation from our shear analysis -- and
to gain an estimate of its magnitude in the autocorrelation case.

A further systematic that is potentially problematic for weak lensing
surveys are color-gradients \citep{voigt12}. However, galaxies
typically have smooth spectral dependence at radio frequencies. This,
combined with the ability to measure the spectral dependence of the
beam very accurately, means that radio weak lensing surveys will be
insensitive to this systematic.

\subsection{Shear measurement}
In order to extract the weak lensing science from the SKA, we will
need to make highly accurate measurements of the shearing of
background sources in real, noisy data. In optical experiments, this
shear measurement process has typically consisted of measuring the
ellipticity of individual galaxies and combining these measurements to
form an estimate of the cosmic shear. A wide variety of galaxy shape
measurement algorithms have been developed for this task and tested
through the STEP and GREAT programmes (see e.g.~\citealt{great3} and
references therein) and experience with real data, honing the
techniques and focusing research on areas which require improvement.
As a consequence, shape measurement from optical data is a
well-established field and techniques already developed are
sufficiently advanced that shape measurement-induced systematics are
likely to be sub-dominant to statistical errors in current and near
future optical lensing surveys. These shape measurement systematics
are often parameterised in terms of the additive bias $c$ and
multiplicative bias $m$ on the ellipticity $e^{obs}$ recovered from an
input source with known true ellipticity $e^{true}$:
\begin{equation}
	e^{obs} = (1 + m)e^{true} + c.
\end{equation}
A formalism is provided in \cite{amara08} for calculating the $m$
and $c$ necessary for systematic errors to be sub-dominant, with
requirements for a selection of SKA surveys shown in
Table~\ref{tab:m_and_c}.
\begin{table}
\centering
\caption{Requirements on multiplicative and additive biases on
  ellipticity measurement for proposed SKA weak lensing surveys to be
  dominated by statistical rather than systematics uncertainties. $Q$
  is a global ``quality factor'' which we calculate 
  from $m$ and $c$ following \cite{voigt10}.}
\label{tab:m_and_c}
\vspace{5mm}
\begin{tabular}{lcccccc}
\hline
Experiment & $A_{sky}$ & $n_{gal}$ & $z_{m}$ & $m<$ & $c<$ & $Q>$ \\
\hline
\hline
SKA1-early & 1000  & 3.0 & 1.0 & 0.014   & 0.0012  & 62 \\
SKA1-early & 5000  & 1.2   & 0.8 & 0.012  & 0.0011  & 79\\
SKA1-early & 30940 & 0.35   & 0.5 & 0.011  & 0.0011 & 80\\
\hline
SKA1       & 1000  & 6.1   & 1.2 & 0.0090   & 0.00095  & 103\\
SKA1       & 5000  & 2.7   & 1.0 & 0.0067  & 0.00082 & 140\\
SKA1       & 30940 & 0.9 & 0.7 & 0.0058  & 0.00076 & 164\\
\hline
SKA2       & 1000  & 37    & 1.6 & 0.0031  & 0.00055 & 318\\
SKA2       & 5000  & 23    & 1.4 & 0.0019  & 0.00043 & 523\\
SKA2       & 30940 & 10    & 1.3 & 0.0012 & 0.00035 & 825\\
\hline
\end{tabular}
\end{table}
However, these algorithms have almost solely been motivated by optical
and NIR data, meaning their robustness to issues peculiar to radio
data, in particular potential systematics from the non-linear
deconvolution process necessary for imaging (and removal of sidelobe
artifacts), is unclear. Of the two radio weak lensing studies so far
performed, both measure shear by modeling data using the orthonormal
shapelets basis set \citep{refregier03}. \cite{patel10} use shapelets
to model images reconstructed using the CLEAN algorithm, while
\cite{chang04} take advantage of the fact that shapelets remain
localised analytic functions under Fourier transformation to directly
model the visibility plane and were able to make a $3.6\sigma$
detection of cosmic shear. More recently, \cite{patel13}
have applied image plane shapelets to simulations, allowing them to
quantify the efficacy of the method, achieving values $m = 0.176$ and
$c = 0.006$ for a simulated e-MERLIN observation. For SKA1, Patel
\emph{et al.} in this volume achieve $m = 0.28$ and $c = 0.001$ for
image plane shapelets, comparable to requirements of $m<0.0054$,
$c<0.0073$ for the SKA1 $5000$ $\mathrm{deg}^2$ survey, as well as
providing an initial comparison of the relative performance of
visibility plane shapelets.

A number of open problems still remain. Although initial steps have
been taken, only a small part of the space of possible radio shear
measurement techniques has so far been explored. Future work may be
expected to take place with two main themes:
\begin{itemize}
	\item Investigation of the ability of image reconstruction
          algorithms to produce images with the level of fidelity
          necessary for weak lensing cosmology and subsequent
          adaptation of known methods from optical/NIR weak lensing to
          radio data. High-fidelity image reconstruction is an active
          research topic in its own right, with many extensions and
          alternatives to traditional CLEAN and Maximum Entropy
          methods under development
          (e.g.~\citealt{sutter14, carrillo14} and references therein).
	\item Investigation of techniques which measure shear directly
          from interferometer visibilities. Such techniques have the
          advantage of avoiding systematics introduced by the
          non-linear deconvolution necessary for the imaging process
          and a visibility plane technique (shapelets in
          \citealt{chang04}) has been the only technique
          so far to successfully detect weak lensing in radio
          data. However, for the datasets produced by the SKA,
          computational challenges for visibility plane techniques
          will be great. Simultaneously fitting $\sim$5 parameter models to
          large numbers of sources over large numbers of visibilities
          is likely to be unfeasible, though averaging and novel
          methods such as $uv$-stacking \citep{lindroos14} have the
          potential to help. There also remains the open question as
          to whether resources to store the raw visibilities from SKA
          observations for later analysis will be available.
\end{itemize}

In the near-term, we expect to begin addressing these issues by
extending previous Gravitational LEnsing Accuracy Testing (GREAT)
challenges (e.g.~\citealt{great3}) to radio data with a
radioGREAT
challenge\footnote{http://radiogreat.jb.man.ac.uk}. This will
aim to better quantify the current status of shape measurement
techniques for radio weak lensing, assess their behaviour across
changing data parameters and spur new interest and developments in the
field.

\subsection{Polarization and rotation velocities as indicators of
  intrinsic alignments}
\label{sec:pol_vrot_techniques}
One unique advantage of radio telescopes for weak lensing is the
polarization information which can provide information on the
intrinsic (unlensed) shapes of background galaxies. As described in
\cite{brown11}, the position angle of the integrated polarized
emission from a background galaxy is unaffected by gravitational
lensing. If the polarized emission (which is polarized synchrotron
emission sourced by the galaxy's magnetic field) is also strongly
correlated with the disk structure of the galaxy then measurements of
the radio polarization position angle can be used as estimates of the
galaxy's intrinsic (unlensed) position angle.

Such an approach could potentially have two key advantages over
traditional weak lensing analyses. Firstly, the polarization technique
can be used to effectively remove the primary astrophysical
contaminant of weak lensing measurements -- intrinsic galaxy
alignments \citep{heavens00, catelan01, hirata04, brown02}
-- which are a severe worry for ongoing and future precision cosmology
experiments based on weak lensing. Secondly, depending
on the polarization properties of distant background disk galaxies,
the polarization technique has the potential to reduce the effects of
noise due to the intrinsic dispersion in galaxy shapes. 

The power of the polarization technique in practice will depend on two
key observational parameters: the scatter in the relationship between
the observed polarization position angle and the intrinsic structural
position angle of the galaxy ($\alpha_{\rm rms}$) and the number of
galaxies for which one can obtain accurate polarization measurements
($n_{\rm pol}$). These parameters depend on the details of the
polarization properties of background galaxies (e.g.~the mean
polarization fraction, $\Pi_{\rm pol}$) which are currently not well
known. There are some existing measurements for a sample of local
spiral galaxies~\citep{stil09} which suggest $\alpha_{\rm rms} <
15^\circ$ and $\Pi_{\rm pol} < 20\%$ although the sample is small. 
Note that it may be possible to select sub-samples of the total galaxy
population to have particular polarization properties. For example,
one could imagine that selecting only galaxies with high fractional
polarization would yield a galaxy sub-sample with highly ordered
magnetic fields which would consequently have a very tight correlation
(low $\alpha_{\rm rms}$) between the polarization orientations and the
intrinsic structural position angles of the galaxies. Of course, such
a sub-sample would also have a very low surface number density of
galaxies. The polarization technique may therefore be better suited to
probing the shear power spectrum on large scales where high number
densities are not required.

A second novel idea that is well suited to radio observations is to
use rotational velocity measurements to provide information about the
intrinsic shapes of galaxies. The idea, first suggested by
\cite{blain02} and \cite{morales06}, is to measure the axis of
rotation of a disk galaxy and to compare this to the orientation of
the major axis of the galaxy disk image. In the absence of lensing,
these two orientations should be perpendicular and measuring the
departure from perpendicularity directly estimates the shear field at
the galaxy's position. Such an analysis would require commensal HI
line observations which could in principle be done at no extra cost in
terms of telescope time. The rotation velocity technique shares many
of the characteristics of the polarization approach described above --
in the limit of perfectly well-behaved disk galaxies, it is also free
of shape noise and it can also be used to remove the contaminating
effect of intrinsic galaxy alignments. In practice, the degree to
which the rotational velocity technique improves on standard methods
will be dependent on observational parameters analogous to the ones
for polarization discussed above. Recently, \cite{huff13} have
proposed to extend this technique using the Tully-Fisher relation to
calibrate the rotational velocity shear measurements and thus reduce
the residual shape noise even further.

Both the polarization technique and the rotation velocity approach are
currently being tested as part of the SuperCLASS, CHILES and CHILES
con Pol projects. They offer great promise for reducing the impact of
shape noise and intrinsic alignments in radio weak lensing surveys.

\section{Weak lensing with HI Intensity Mapping}
\label{sec:HI}
Weak gravitational lensing of high redshift 21cm emission has been the
subject of several studies focusing on the Epoch of Reionization
(EoR) observations. In \cite{Zahn:2005ap} and \cite{Metcalf:2009} it
was shown that if the EoR is at redshift $z\sim8$ or later, a large
telescope like the SKA could measure the lensing power spectrum and
constrain the standard cosmological parameters. The authors extended
the Fourier-space quadratic estimator technique, which was first
developed in \cite{Hu:2001tn} for CMB lensing observations to three
dimensional observables, i.e. the 21 cm intensity field
$I(\theta,z)$. These studies did not consider 21 cm observations from
redshifts after reionization when the average HI density in the
Universe is much smaller.

HI intensity mapping is a technique that has been proposed for
measuring the distribution of HI gas before and during reionization
(see Pritchard \emph{et al.} in this volume) and
measuring the BAO at redshifts of order unity \citep{Chang:2008,
  Chang:2010, Seo:2009fq, Masui:2010, Ansari2012, Battye:2012tg,
  Chen2012, Pober2013}. In this technique, no attempt is made to
detect individual objects. Instead the 21 cm emission is treated as a
continuous three dimensional field. The angular resolution of the
telescope need not be high enough to resolve individual galaxies which
makes observations at high redshift possible with a reasonably sized
telescope.

A recent study \citep{Pourtsidou:2014} extended the
21cm-intensity-mapping-lensing method further, taking into account the
discreteness of galaxies, and investigated the possibility of
measuring lensing at intermediate redshifts without resolving (in
angular resolution) or even identifying individual sources. Here we
present an improved analysis of the signal-to-noise expected from an
SKA-like telescope array, and demonstrate the performance of the
SKA-Mid facility for SKA1 and for SKA2. 

One of the first objectives of a $21\rm{ cm}$ lensing survey will be
to measure the two-point statistics of the convergence field
$\kappa(\vec{L},z_s)$ or, equivalently, the displacement field $\delta
\theta(\vec{L},z_s)$, averaged over $z_s$. That is,
\begin{equation}
C_L^{\delta \theta \delta \theta} = \frac{9 \Omega^2_m H^3_0}{L(L+1)
  c^3} \int_0^{z_s}dz \, P(k=L/r(z),z) [W(z)]^2 /E(z),
\end{equation} 
where $W(z)=(r(z_s)-r(z))/r(z_s)$, $E(z)=H(z)/H_0$. The expected error
in the power spectrum $C_L^{\delta \theta \delta \theta}$ averaging
over $\vec{L}$ directions in a band of width $\Delta L$ is given by
\begin{equation}
\label{eq:DCL}
\Delta C_L^{\delta \theta \delta \theta} =\sqrt{\frac{2}{(2L+1)\Delta
    L f_{\rm{sky}}}}\left(C_L^{\delta \theta \delta
  \theta}+N_L\right).
\end{equation} 

Here, $N_L$ is the derived $\delta \theta$ estimate reconstruction
noise which involves the underlying dark matter power spectrum, the HI
density $\Omega_{\rm HI}(z)$ as well as the HI mass moments up to 4th
order and, of course, the thermal noise of the telescope (see
\citealt{Pourtsidou:2014} for details). For the HI mass function we will
first assume the most conservative scenario, i.e. a no evolution model
using the results from the HIPASS survey \citep{Zwaan03}, giving
$\Omega_{\rm HI} \simeq 4.9 \times 10^{-4}$ (Model A). We will also
consider a more realistic evolution scenario from \cite{Zhang06} where
$\Omega_{\rm HI}(z) = 4.9 \times 10^{-4} \,(1+z)^{2.9} {\rm
  exp}(-z/1.3)$.  This is expressed through an evolution of the
normalization of the Schechter function $\phi^\star$ (Model B).

\begin{figure}[h]
\vspace{-1cm}
\centerline{
\includegraphics[scale=0.3]{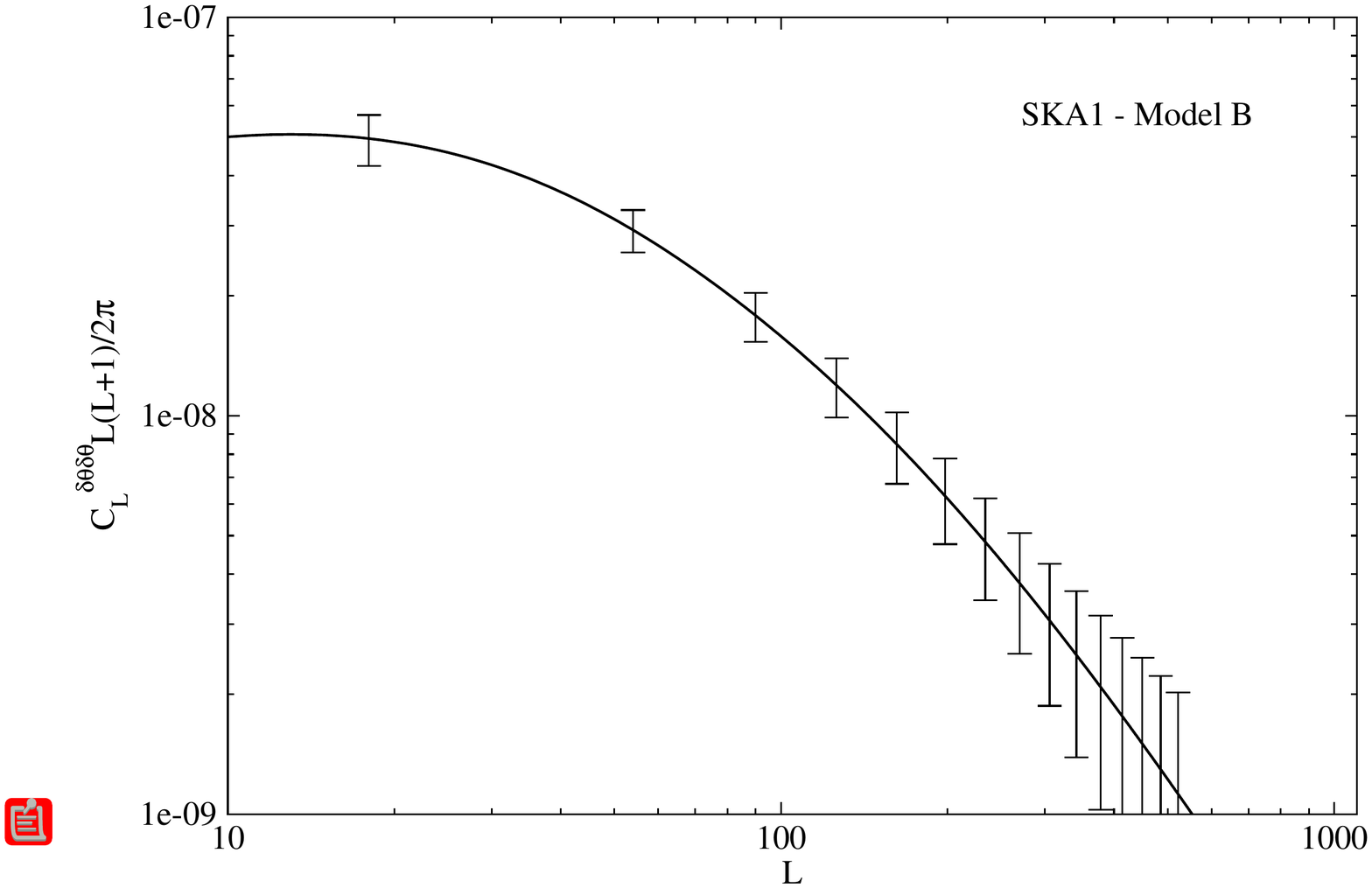}
\includegraphics[scale=0.3]{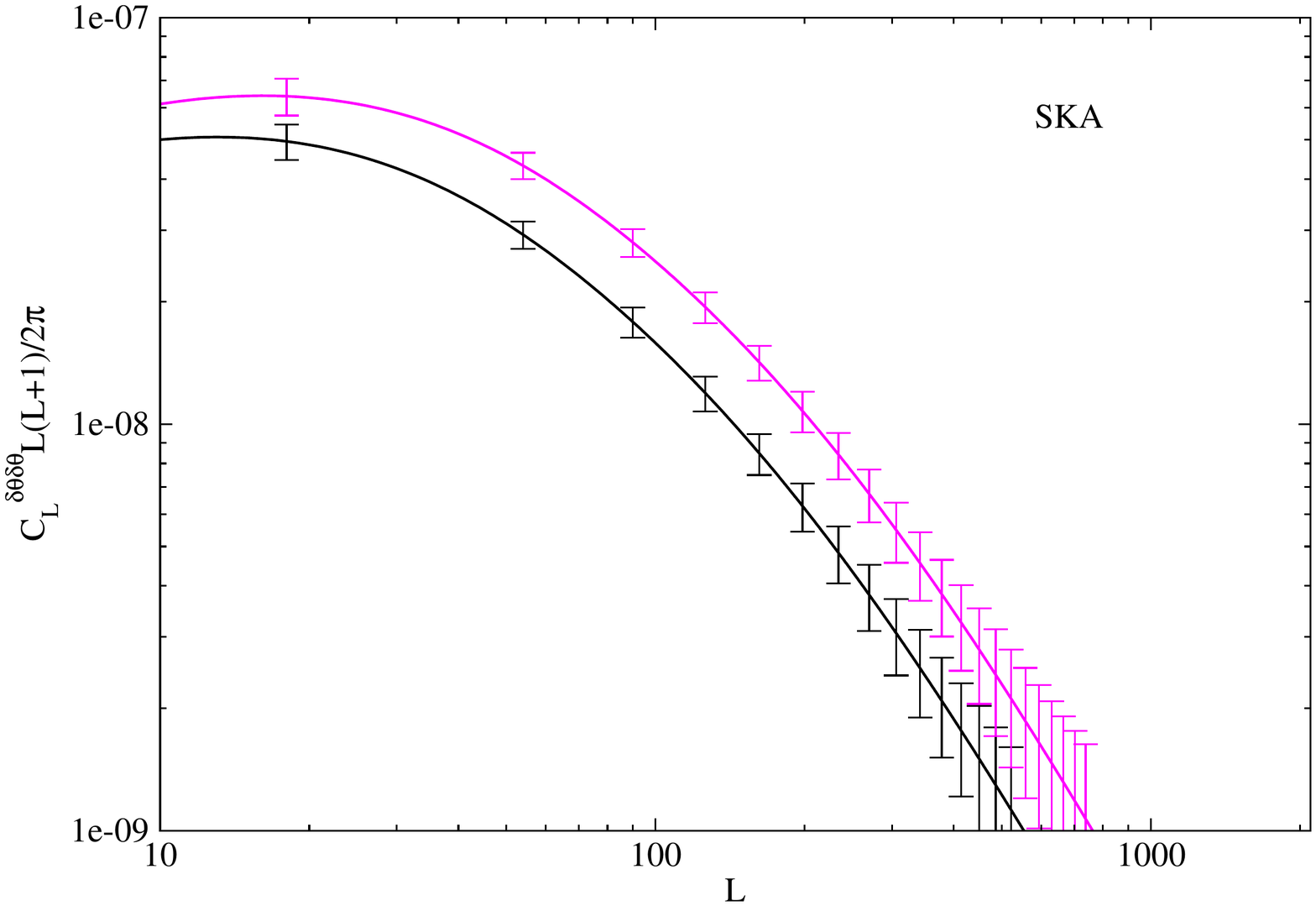}}
\caption{{\it Left}: Displacement field power spectrum for $z_s=2$ and the
  corresponding measurement errors using the SKA1-Mid specifications
  and Model B for the HI mass function. {\it Right}: Displacement field
  power spectrum for $z_s=2$ (solid black line) and $z_s=3$ (solid
  magenta line) and the corresponding measurement errors using the
  SKA2 specifications and Model A (no evolution) for the HI mass
  function. See the main text of Section~\protect\ref{sec:HI} for
  details of the HI mass function adopted for Models A and B.}
\label{fig:SoverN}
\end{figure}

Here we will concentrate on source redshift $z_s \sim 2$.  The
SKA1-Mid array operates in the frequency range $\sim 350-3050 \, {\rm
  MHz}$ divided in three bands \citep{Dewdney:2013}. Our chosen $z_s$
corresponds to frequency $\nu_s=473 \, {\rm MHz}$ (Band 1).  The power
spectrum of the thermal noise of a SKA-like interferometer core array
(using the uniform distribution approximation) is $ C^{\rm N}_\ell =
\frac{(2\pi)^3 T^2_{\rm sys}}{B t_{\rm obs} f^2_{\rm cover} \ell_{\rm
    max}(\nu)^2} \, , $ where the system temperature $T_{\rm
  sys}=(40+1.1T_{\rm sky}) \, {\rm K}$ with $T_{\rm sky}=60
\lambda^{-2.55}$, $B$ is the chosen frequency window, $t_{\rm obs}$
the total observation time, $D_{\rm tel}$ the diameter (maximum
baseline) of the array, $\ell_{\rm max}(\lambda)=2\pi D_{\rm
  tel}/\lambda$ is the highest multipole that can be measured by the
array at frequency $\nu$ (wavelength $\lambda$), and $f_{\rm cover}$
is the total collecting area of the telescopes $A_{\rm coll}$ divided
by $\pi(D_{\rm tel}/2)^2$. For SKA-Mid we can consider a $2 \, {\rm
  yr}$ observation time, $f_{\rm sky}\sim 0.7$ (corresponding to $\sim
30,900 \, {\rm deg}^2$ survey area), and we further choose $B = 20 \,
{\rm MHz}$ and $\Delta L=36$. Keeping these values constant, in
Fig.~\ref{fig:SoverN} we present forecasts for the displacement field
power spectrum for source redshifts $z_s=2$ and $z_s=3$. For the Model
A HI mass function, SKA1 does not have the sensitivity to measure the
signal. However, for Model B we see a great improvement in the
signal-to-noise ratio, such that the lensing signal can be well
measured by SKA1 (Fig.~\ref{fig:SoverN}, left panel). For SKA2, a
precise characterization of the power spectrum is possible at multiple
redshifts even in the conservation no-evolution scenario (Model A,
Fig.~\ref{fig:SoverN}, right panel).

This technique should enable us to measure the lensing power spectrum
at source redshifts well beyond those accessible with more traditional
weak lensing surveys based on the shearing of individual galaxy
images. This is a promising approach to investigating the possible
evolution of dark energy or modified gravity at high redshift. In
addition to measuring the power spectrum, the SKA-Low array should be
able to map the dark matter distribution with high fidelity in a
5-by-5 degree field using this technique (see Pritchard \emph{et al.} in this
volume for details).

\section{Galaxy and cluster weak lensing}
In addition to probing large-scale structure, weak lensing can be used
to measure the distribution of mass (both dark and bright) around
individual galaxies, as well as other structures such as galaxy pairs,
groups and clusters, or
voids~\citep{mandelbaum06b,gillis13,melchior14}. Since the signal from
an individual lens is too weak to be detectable, except for massive
clusters, such ``galaxy-galaxy'' measurements are usually stacked to
produce an average mass distribution around a certain type of lens
objects.  On large, linear scales galaxy-galaxy lensing constrains
galaxy bias (e.g.~\citealt{hoekstra02}). Moreover, interpreted as a
galaxy position-shear cross-correlation it can provide additional
cosmological constraints and valuable calibration of astrophysical
systematics~\citep{bernstein09, joachimi10} and Kirk \emph{et
  al.}, this volume).  On smaller scales galaxy-galaxy lensing is
sensitive to the radial matter density profile of galaxy haloes as
well as the abundance and distribution of substructure in its
outskirts, all as a function of galaxy properties such as type,
stellar mass, or velocity dispersion~\citep{vanUitert13}.

The SKA surveys will provide a unique angle to galaxy-galaxy weak lensing
studies, with key contributions in the following areas:
\begin{itemize}
\item \textbf{Competitive or superior statistical errors}: These
  errors are driven by the number density of both lens and source
  galaxies (particularly on small scales) and the sample variance due
  to a finite sky coverage (particularly on large scales).
\item \textbf{Spectroscopic redshifts for new galaxy samples}: To
  obtain unbiased measurements, a clean separation of lens and source
  samples is paramount. Spectroscopic redshifts for foreground objects
  allow for a clean and precise definition of lens samples and lens
  properties. Accurate redshifts for a subset of the source sample
  characterise the redshift distribution of background objects.
\item \textbf{Novel lens galaxy sample definitions}: The information
  from radio wavelengths can be used to define new lens samples
  e.g. in terms of their radio luminosity, AGN activity, or gas
  abundance. This allows for the measurement of the relation between
  these properties and matter halo characteristics, e.g. the HI gas
  mass to halo mass ratio as a function of redshift.
\end{itemize}

\begin{figure}[t!]
\centering
\includegraphics[scale=.38, trim = 4.5cm 0.0cm 0.0cm 0.0cm, angle=270]{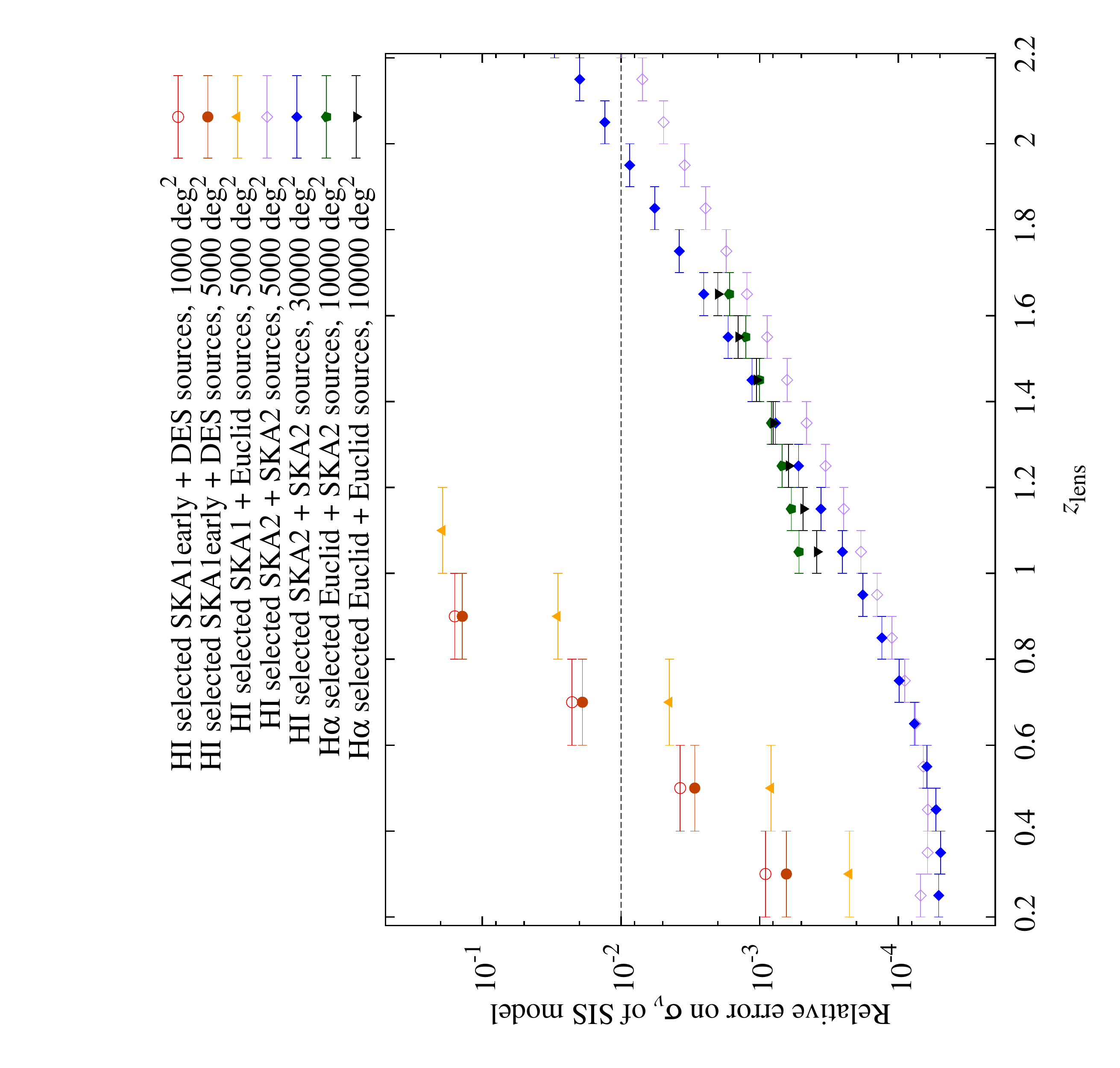}
\caption{Relative $1\sigma$ error on the velocity dispersion,
  $\sigma_v$, of a singular isothermal sphere (SIS) profile as a
  function of the redshift of the lensed objects, constrained via the
  average tangential shear signal in the transverse separation range
  $r_p \in [0.1;1]\,$Mpc/$h$. The legend lists the lens sample
  selection, the source sample from which the shear signal is
  measured, and the assumed overlap in survey area. The error bars
  show the assumed width of the redshift bins. A $1\,\%$ constraint is
  indicated by the horizontal dashed line.}
\label{fig:sisconstraints}
\end{figure}

To quantify this performance, we forecast expected errors on a
one-parameter radial matter profile constrained via the average
tangential shear measured in the range of 0.1 Mpc/$h$ to 1 Mpc/$h$
transverse separation from the lenses. We use a singular isothermal
sphere (SIS) as a simple profile, 
\eq{ \rho_{\rm SIS}(r) = \frac{\sigma_v^2}{2 \pi G r^2}\;, }
where $r$ denotes three-dimensional distance, and where $\sigma_v$ is
the velocity dispersion, which is the single free parameter. To
facilitate comparison, we set $\sigma_v=250$ km/s for all
calculations. The tangential shear signal for this profile is given by
(assuming a spatially flat Universe, \citealt{bartelmann01}) 
\eq{
  \ba{\gamma_+}(r_p) = 2 \pi \br{\frac{\sigma_v}{c}}^2\;
  \frac{\chi_{\rm L}}{r_p}\; \br{1 - \frac{\chi_{\rm L}}{\chi_{\rm
        S}}}\;, } 
where $\chi_{\rm L}$ is the comoving distance to the lens, and
$\chi_{\rm S}$ the comoving distance to the source galaxy used to
estimate the gravitational shear. 

In Fig.~\ref{fig:sisconstraints} we show the forecasted relative error
on the SIS free parameter $\sigma_v$ for a selection of lens and
source sample combinations. We have divided the lens samples into
redshift bins of width 0.1 and 0.2, respectively.  It is possible to
explore new terrain in galaxy-galaxy lensing already with early SKA1
data by selecting lens samples using spectroscopic redshifts from an
HI survey. Combining this with a source sample from an optical survey
such as the Dark Energy Survey (DES), one can obtain percent-level
constraints out to $z \sim 0.6$ for a few subsamples (e.g.~selected on
gas abundance). The full SKA1 in combination with Euclid imaging for a
source sample will decrease errors further by about a factor of five
and extend the range where lenses can be studied to $z=0.8$.

SKA2 will enable unique and powerful galaxy-galaxy lensing
measurements without the need to rely on optical counterparts. As
Fig.~\ref{fig:sisconstraints} shows, it will generate an unprecedented
redshift baseline to study halo properties with a single facility. Out
to $z \sim 1.5$ constraints are below the per mil level, which allow
for tight constraints on complex models and/or the division into
several subsamples. SKA2 will also be on par with Euclid in measuring
the galaxy-galaxy lensing signal of emission-line (mostly H$\alpha$)
galaxies detected in the Euclid spectroscopic survey, providing
independent corroboration for this high-redshift measurement. Note
finally that the very deep SKA2 surveys will allow the application of
the novel techniques described in
Section~\ref{sec:pol_vrot_techniques} as well as magnification
analyses (e.g.~\citealt{morrison12}) in a galaxy-galaxy lensing context.

\section{Concluding remarks}
The field of weak lensing with radio surveys is currently in its
infancy but the prospect of SKA surveys coming online at the end of
this decade presents a huge opportunity for opening a new wavelength
window on the dark Universe. We have argued in this Chapter that the
radio band and the SKA in particular offer exciting prospects for
enhancing and extending the reach of weak lensing studies beyond the
limits of what is possible using their traditional optical
approaches. 

The SKA surveys will probe the galaxy population to higher redshift
where the lensing signal is larger and potentially easier to
measure. There is huge potential to exploit the cross-correlation of
lensing shear measurements obtained from overlapping optical and radio
surveys which can be powerful in mitigating a number of systematic
effects. 

Moreover, the radio band offers the exciting possibility of applying
new techniques to measure weak lensing signals. One can make use radio
polarization and/or rotational velocity information (from HI
observations) to mitigate against intrinsic galaxy alignments which
are the most worrying systematics for future precision weak lensing
measurements. The opportunity to analyze HI Intensity Mapping surveys
for the effects of weak lensing will open a new window on the dark
Universe at redshifts that are well beyond the reach of traditional
cosmic shear techniques.

The SKA also offers a new route to performing galaxy-galaxy
lensing where spectroscopic redshifts from HI observations and the
extra diagnostics coming from the radio band will allow for the
precise definition of new lens samples.

As with other future large-scale survey instruments, weak lensing
studies with the SKA will facilitate the auto- and cross-correlation
of galaxy clustering and cosmic shear measurements within the same
experimental set up. Generally speaking, the joint exploitation of
clustering and lensing is extremely useful as the two probes are
highly complementary. Constraints from baryonic acoustic oscillations
and/or redshift space distortions are ``orthogonal'' to weak lensing,
especially in the presence of uncertainties in photometric redshifts
and inaccurate knowledge of galaxy clustering bias \citep{zhan06,
  abate12, laureijs11, amendola13}.

In terms of constraining power for dark energy and/or modified
gravity, galaxy clustering data measure the ``dark fluid'' equation of
state at higher redshift than other standard candles like type Ia
supernovae. Furthermore, clustering probes the evolution of matter
fluctuations, which means that, through the Poisson equation, it is a
proxy for the Newtonian potential $\Phi$.  In the case of modified
gravity, clustering is sensitive to modifications occurring to the
Newtonian gravitational constant. On the other hand, the deflection
potential is proportional to the sum of the two metric potentials --
which are equal in GR but not in more general gravity
theories. Through this dependence, lensing effects represent a direct
probe of non-standard gravity behaviours. As a consequence, the
sensitivity to beyond-GR growth parameters comes in its greatest part
from weak lensing, which provides the only direct measurements of
growth (without biasing), e.g.~\cite{weinberg13}.

In order to realise the significant and exciting potential that radio
weak lensing has to offer, a great deal of development work on
analysis techniques, and demonstrating those techniques on real radio
data from precursor surveys will be required. A key challenge will be
to develop the field of galaxy shape estimation for radio observations
to the level of maturity that is currently enjoyed by the optical
lensing community. Work has already started in this direction and will
gain further momentum through the ongoing ``radioGREAT'' challenge
during the coming  year. At the same time, surveys with e-MERLIN,
LOFAR and the JVLA are starting to produce the high-resolution,
wide-field data suitable for demonstrating radio weak lensing
techniques on real observations. The planned program of algorithm
development work coupled with the successful application of the
developed radio lensing tools to the SKA pathfinder and precursor
surveys will position the community favourable to fully exploit the
tremendous opportunity for radio weak lensing that will be afforded by
the commissioning of the SKA at the end of this decade.

\section*{Acknowledgments}
MLB is a STFC Advanced/Halliday fellow. MLB, SC and IH
are supported by an ERC Starting Grant (grant no. 280127). DJB is
supported by UK Science and Technologies Facilities grant,
ST/K00090X/1. PP is funded by a SKA South Africa Postdoctoral
Fellowship.

\bibliographystyle{apj}

\end{document}